# Voltage-controlled antiferromagnetism in magnetic tunnel junctions

Meng Xu[1], Mingen Li[2], Pravin Khanal[1], Ali Habiboglu[1], Blake Insana[1],
Yuzan Xiong[3, 4], Thomas Peterson[5], Jason C. Myers[6], Deborah Ortega[1],
Hongwei Qu[4], C.L. Chien[2], Wei Zhang[3] Jian-Ping Wang[5] and W.G. Wang[1]*

1. Department of Physics, University of Arizona, Tucson, Arizona 85721, USA
2. Department of Physics and Astronomy, The Johns Hopkins University, Baltimore, Maryland 21218, USA
3. Department of Physics, Oakland University, Rochester, Michigan 48309, USA
4. Department of Electrical and Computer Engineering, Oakland University, Rochester, Michigan 48309, USA
5. Department of Electrical and Computer Engineering, University of Minnesota, 200 Union Street SE, Minneapolis, Minnesota 55455, USA
6. Characterization Facility, University of Minnesota, 100 Union Street SE, Minneapolis, Minnesota 55455, USA

**Abstract**: We demonstrate a voltage-controlled exchange bias effect in CoFeB/MgO/CoFeB magnetic tunnel junctions that is related to the interfacial $Fe(Co)O_x$ formed between the CoFeB electrodes and the MgO barrier. The unique combination of interfacial antiferromagnetism, giant tunneling magnetoresistance, and sharp switching of the perpendicularly-magnetized CoFeB allows sensitive detection of the exchange bias. It is found that the exchange bias field can be isothermally controlled by magnetic fields at low temperatures. More importantly, the exchange bias can also be effectively manipulated by the electric field applied to the MgO barrier due to the voltage-controlled antiferromagnetic anisotropy in this system.

*wgwang@physics.arizona.edu

Recent research on the interaction between electric fields and magnetic order has yielded some very interesting results [1–4]. In ferromagnetic systems with 3$d$ transitional metals, it was shown that electric fields can effectively change the perpendicular magnetic anisotropy (PMA) [5–7], leading to low energy switching of magnetization on the sub-ns time scale [8,9]. Electric fields driven ionic migration can also have a large impact on the saturation magnetization of the 3$d$ ferromagnets (FM), resulting in a giant modulation of magnetism [10–12]. On the other hand, antiferromagnets (AFM) possess several unique features, such as staggered arrangement of spins, resilience to external magnetic fields, and much higher resonance frequency in the THz range [13–15]. Electrical control of magnetism in antiferromagnetic (AF) systems has also attracted a great deal of attention. A pioneering example can be found in $Cr_2O_3$ [16]. The AF order parameter of the magnetoelectric $Cr_2O_3$ can be reversibly and deterministically switched by voltages, manifested by the exchange bias effect experienced by the adjacent FM layer [17–19], similar to those observed in multiferroic systems [20,21]. Other examples include noncentrosymmetric antiferromagnets, such as $CuMnAs$ [22] and $Mn_2Au$ [23], where the AF order parameter may be switched by electrical currents. The AF order can also be electrically controlled by adjusting exchange spring [24], by spin-orbit torques [25], and through coupling with a ferroelectric substrate as recently demonstrated [26].

In this work, we report the discovery of the voltage-controlled antiferromagnetism in a model spintronics system that is also technologically important: the CoFeB/MgO/CoFeB perpendicular magnetic tunnel junction (pMTJ). It is demonstrated for the first time through transport measurement that at low temperature the CoFeB layers are exchange-biased to the interfacial Fe(Co)$O_x$ layers naturally formed between CoFeB and the MgO barrier. In addition to the voltage-controlled magnetic anisotropy (VCMA) effect [5–7], the electric field can have a profound impact on the exchange bias field ($H_{EB}$) in this system. Thanks to the giant tunneling magnetoresistance (TMR) and the sharp switching of the CoFeB layers with PMA, a clear dependence of $H_{EB}$ on the applied voltage has been detected, which is attributed to the voltage control of AFM anisotropy.

The MTJ thin films are fabricated using magnetron sputtering with the core structure $Co_{20}Fe_{60}B_{20}$(0.8 nm)/MgO(1-3.5 nm)/$Co_{20}Fe_{60}B_{20}$(1.6 nm). More information on the MTJ fabrication can be found in the Supplementary Material [27]. After annealed at 300 °C for 10 minutes, the TMR curve of a junction was measured under perpendicular field at room temperature (RT) as shown in Fig. 1a. The barrier thickness is 3 nm and the resistance area product (RA) is $5 \times 10^7$ $\Omega \mu m^2$. The black curve exhibits sharp switches with a TMR value of 120%, a ratio that is typical for MTJs annealed at this condition [30]. Then the sample is cooled under a perpendicular field of +3000 Oe. The resistance of the antiparallel state ($R_{AP}$) consistently increases with decreasing temperature, while the resistance of the parallel state ($R_P$) remains largely unchanged, leading to a TMR value that is almost doubled compared with that at RT (Supplementary Fig. S1). Unlike the TMR curve measured at RT, where the switching fields of the soft (bottom, 0.8 nm) and hard (top, 1.6 nm) CoFeB layers are symmetric about zero field, the switching fields at 30 K are shifted negatively. Reversing the cooling field to -3000 Oe changes the switching fields to the positive direction as shown by the light blue curve. This is direct evidence of the exchange bias effect in the system. Here we define the exchange bias field $H_{EB}$ as $(H_{C1} + H_{C2})/2,$ where $H_{C1}$ ($H_{C2}$) is the negative (positive) switching field of a given FM layer. The magnitude of $H_{EB}$ is about 100 Oe for the soft layer and 350 Oe for the hard layer with both polarities opposite to the cooling field.

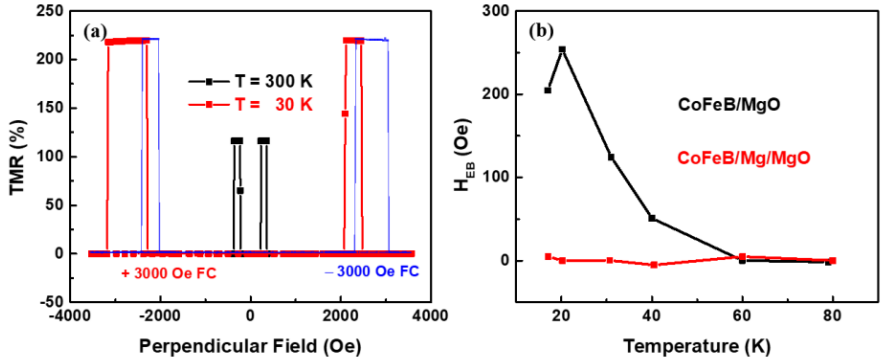

**Fig. 1** (a), TMR curves of a pMTJ measured at RT (black) and 30 K (red and light blue). While no exchange bias is observed at RT, after cooling down to 30 K under a perpendicular field of ±3000 Oe, obvious $H_{EB}$ for the soft and hard FM layers can be observed. (b), Temperature dependence of $H_{EB}$ measured for 2 samples in the Hall bar geometry, showing the observed exchange bias originates from the FM/MgO interface.

To the best of our knowledge, this is the first observation of exchange bias via the TMR measurement in a pMTJ with no active AF layers such as IrMn or PtMn. Previously the interfacial spins of a thick Fe layer (10nm) adjacent to MgO were demonstrated to be exchange-biased using magnetization-induced second harmonic generation (MSHG) technique [42], and the existence of exchange bias was indirectly inferred by the FMR measurement [43]. Here due to the reduced thickness of the FM layers (~1nm), the entire magnetization, instead of only the interfacial spins [42], is exchange-biased. With the sharp switching of perpendicularly-magnetized CoFeB, now the exchange bias effect can be sensitively detected by the TMR measurement. It is known that oxygen ions generated during RF sputtering can bombard and oxidize adjacent Fe(Co) to form $FeO_x$ and $CoO_x$ at the bottom CoFeB/MgO interface, while thermally induced oxidation/reduction could also take place during the post-growth annealing [41]. Therefore, the exchange bias effect is most likely linked to the antiferromagnetism associated with the Fe (Co)-O bonding. To verify this assumption, two Hall bars resembling the bottom FM layer of the pMTJs are fabricated, one with the structure of Ta/CoFeB(0.8 nm)/MgO(3 nm)/Ta and the other with the structure of Ta/CoFeB(0.8 nm)/Mg(0.6 nm)/MgO(3 nm)/Ta. It is expected that the Fe (Co)-O bonding in the latter is greatly weakened due to the Mg insertion. As illustrated by the temperature dependence of $H_{EB}$ in Fig. 1b, the exchange field in the sample with Mg insertion is dramatically reduced, thus unambiguously demonstrating that the exchange bias effect observed in Fig. 1a is caused by the antiferromagnetism associated with Fe (Co)-O bonding. The pMTJs in this study are of high quality, as demonstrated by the large TMR ratio close to 400% at low temperature (Supplementary Material), indicating no severe oxidation of the CoFeB layers. Therefore the $FeO_x$ or $CoO_x$ formed between CoFeB and MgO is likely to be only a monolayer or sub-monolayer, which is consistent with the sub-100 K blocking temperature ($T_B$) observed.

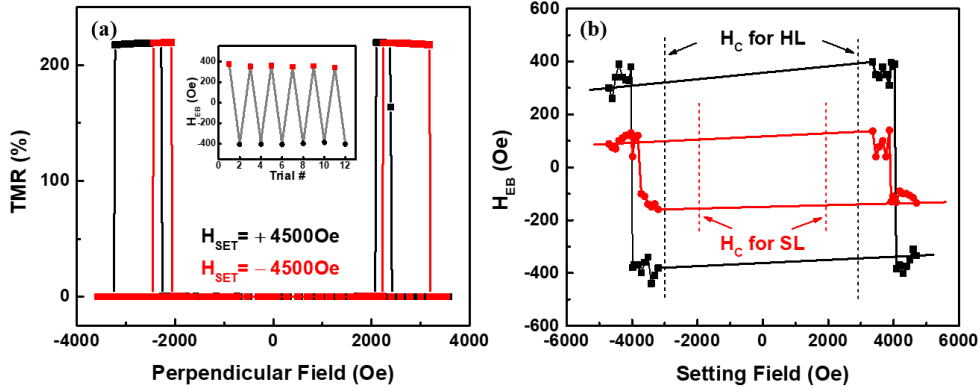

**Fig. 2** (a), Representative TMR curves taken after the application of -4500 Oe setting field (red), and after the application of +4500 Oe setting field (black). Inset show the reversible and deterministic control of $H_{EB}$ by applying setting field isothermally. (b), $H_{EB}$ dependence on external setting field for both FM layers. Dashed lines indicated their respective coercivities. All data are collected at the constant temperature of 30K.

To this point, the exchange bias effect in our sample behaves similarly to those in a typical FM/AFM bilayer system [44]. Generally speaking, the exchange bias field is only observed when the FM is cooled below $T_N$ under an external magnetic field. After the $H_{EB}$ has been set, its polarity cannot be changed solely by reversing the magnetic field without warming above $T_N$. In rare cases, isothermal switching of $H_{EB}$ could happen under spin flop transition, but with very high switching fields (over 10 T) [45]. In sharp contrast, isothermal switching of $H_{EB}$ can be achieved in our system with a relatively small magnetic field as demonstrated in Fig. 2a. First, the pMTJ was cooled from RT to 30K under a positive magnetic field, thereby setting a negative $H_{EB}$ for both FM layers. Then a setting field of -4500 Oe was applied isothermally, followed by the subsequent measurement of the TMR curve. Interestingly, positive $H_{EB}$ values of +100 Oe and + 350 Oe were obtained for the two FM layers. The same procedure was performed with + 4500 Oe setting field and the $H_{EB}$ switched to -100 Oe and -350 Oe. The relatively small setting field of ± 4500 Oe indicates the antiferromagnetism at the FM/oxide interface is weak, which is consistent with the very thin $Fe(Co)O_x$ formed between the CoFeB and MgO. Notably, the $H_{EB}$ can be reversibly and deterministically controlled by the setting field as shown in the inset of Fig. 2a, demonstrating the robustness of this effect. Here we define the minimum field that switches $H_{EB}$ isothermally as $H_{AF}$. To determine $H_{AF}$, the sample was first initialized with a large positive (negative) field (> 6000 Oe), which set the $H_{EB}$ to be negative (positive). Then the TMR curves were measured after incremental setting field applied and the corresponding $H_{EB}$ values are plotted in Fig. 2b. Two features immediately emerge in the setting field dependence of $H_{EB}$. First, $H_{AF}$ is symmetric about the zero setting field. Second, $H_{AF}$ is different for soft and hard layers, which further confirms the interfacial origin of the exchange bias. Note the exchange bias can only be observed when the maximum magnetic field applied during TMR measurement is smaller than $H_{AF}$. Otherwise, the exchange bias will simply manifest itself as coercivity enhancement and there will be no shift in the TMR curve.

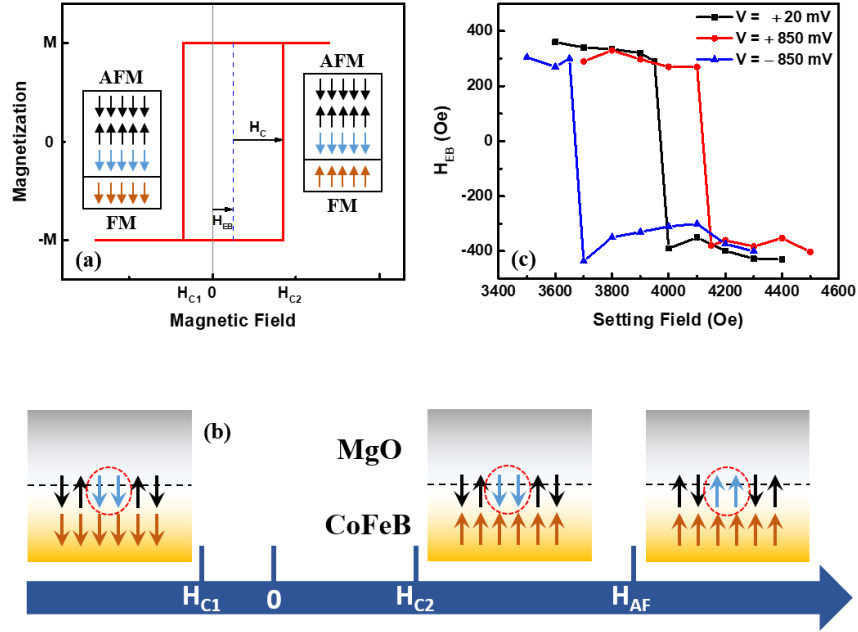

**Fig. 3** (a), Schematic picture of the exchange bias effect in AFM/FM bilayers, where the exchange coupling between the FM spins (brown) and the uncompensated interfacial spins (blue) of the AFM (black) leads to a shift of the hysteresis loop characterized by $H_{EB}$. (b), Schematic diagram of the isothermal control of $H_{EB}$ with magnetic field. $H_{C1}$ and $H_{C2}$ are the magnetization switching fields in negative and positive field direction. $H_{AF}$ is the critical setting field above which $H_{EB}$ polarity changes. (c), Setting field dependence of the hard layer $H_{EB}$ in a pMTJ under different voltages, in which the positive voltage increases $H_{AF}$ and the negative voltage reduces $H_{AF}$.

The exchange bias effect can be usually understood through the schematic diagram shown in Fig. 3a [44]. The bottom FM spins are coupled to the uncompensated interfacial spins (blue arrows) that are strongly pinned by the AFM (black arrows). Since the AFM is insensitive to the external field, FM spins are biased to one specific direction determined by the direction of uncompensated spins and the nature of exchange coupling (FM or AF). Within this context, the exchange bias effect in our sample can be schematically represented by a simplified model shown in Fig. 3b. The interfacial $Fe(Co)O_x$ layer serves as the AF layer (black arrows) between CoFeB and MgO. The circled spins are uncompensated and moderately pinned by the surrounding AF spins. For simplicity, the AF spins are shown to form one monolayer. In reality, they may form discrete patches that are more or less randomly distributed at the interface [42], as indicated by the spin-glass like feature shown in Fig. S2 of the Supplementary Material. The isothermal control of $H_{EB}$ by the magnetic field stems from the weak AF order of the interfacial $Fe(Co)O_x$. When the external field exceeds $H_{C2}$ but below $H_{AF}$, the circled (pinned) spins and the surrounding AF spins remain unchanged, leading to an $H_{EB}$ observed in the hysteresis loop. When the external setting field exceeds $H_{AF}$, the pinned spins would be reversed to lower the Zeeman energy, which simultaneously rotates the AF spins. Subsequent TMR measurement would produce an $H_{EB}$ with the opposite sign. This picture qualitatively describes the isothermal switching of $H_{EB}$ presented in Figure 2. $H_{EB}$ appears when $H_C < H_{AF}$ and the applied magnetic field is smaller than $H_{AF}$.

One important feature of the CoFeB-MgO pMTJ is the interfacial PMA and the associated VCMA effect. The PMA is the result of the hybridization of 3d orbitals of Fe and 2p orbital of Oxygen [46]. When an electric field is applied to the junction, it modifies the electron occupation in different orbitals through the Fe-O bonding, thus leading to a change of magnetic anisotropy energy [47]. Since the presence of exchange bias and isothermal control of $H_{EB}$ by the magnetic field are also caused by Fe(Co)-O bonding at the interface, one may anticipate a correlation between the exchange bias and the electric field in the system. To test this point, an experiment similar to the one in Fig. 2b was performed, only this time with voltage applied to the junction when the setting magnetic field is turned on. Note the $H_{EB}$ is still measured by the TMR curve under low bias voltage (~10 mV) after the removal of the setting field and voltage. Remarkably, a clear dependence of $H_{AF}$ on the applied voltage now appears. As shown in Fig. 3c, $H_{AF}$ can be effectively increased (reduced) by positive (negative) voltages. The applied voltage of ± 850 mV can cause a total modification of 400 Oe in $H_{AF}$. Note only the positive branch of the hard layer $H_{EB}$ is plotted in Fig. 3c to highlight the changes brought on by voltage.

If the voltage can modulate $H_{AF}$, one may expect the associated change of antiferromagnetism to be reflected by the TMR curve as well. The magnetoresistance of the pMTJ was measured at different voltages and plotted in Fig. 4a. Indeed, very different TMR curves are obtained. It is known that the $H_C$ of the two FM layers can be altered by the VCMA effect, which gives rise to very distinct TMR curves under different voltages [7]. The pMTJs in this study do exhibit the conventional VCMA effect, as evidenced by the voltage dependence of the switching fields for the hard layer at RT (Fig. 4b), where modulation of coercivity is centered around zero field. The behavior of TMR curves in Fig 4a, however, are markedly different from those influenced only by the VCMA effect. Here the pronounced changes are observed under *positive* magnetic field only. A closer inspection reveals that now the changes of $H_{C1}$ and $H_{C2}$ are asymmetric about zero magnetic field as shown in Fig. 4c. This sample was initialized with a large negative field, resulting in a positive $H_{EB}$. As positive voltage leads to a stronger interaction between the pinned uncompensated spins and the surrounding AF regions (Fig. 3c), the FM magnetization becomes harder to switch up (leading to a larger $H_{C2}$) and easier to switch down (leading a smaller $H_{C1}$). Under negative magnetic field, this phenomenon counteracts with the VCMA effect that enhances $H_{C1}$, leading to a weaker dependence of $H_{C1}$ on voltage. Simultaneously, a stronger voltage dependence of $H_{C2}$ emerges as these two effects facilitate each other under positive magnetic field. A direct result of this asymmetric dependence of the switching fields on voltage is the shift of the hysteresis loop by voltage, namely the voltage-controlled exchange bias effect. The voltage dependence of $H_{EB}$ is plotted in Fig. 4d. A monotonic (but nonlinear) dependence of $H_{EB}$ on voltage can be seen, where the exchange bias is enhanced by positive voltage and reduced by negative voltage. The modification of $H_{EB}$ is quite significant, with a change of more than 150 Oe observed when the voltage is varied between +1V and -1V.

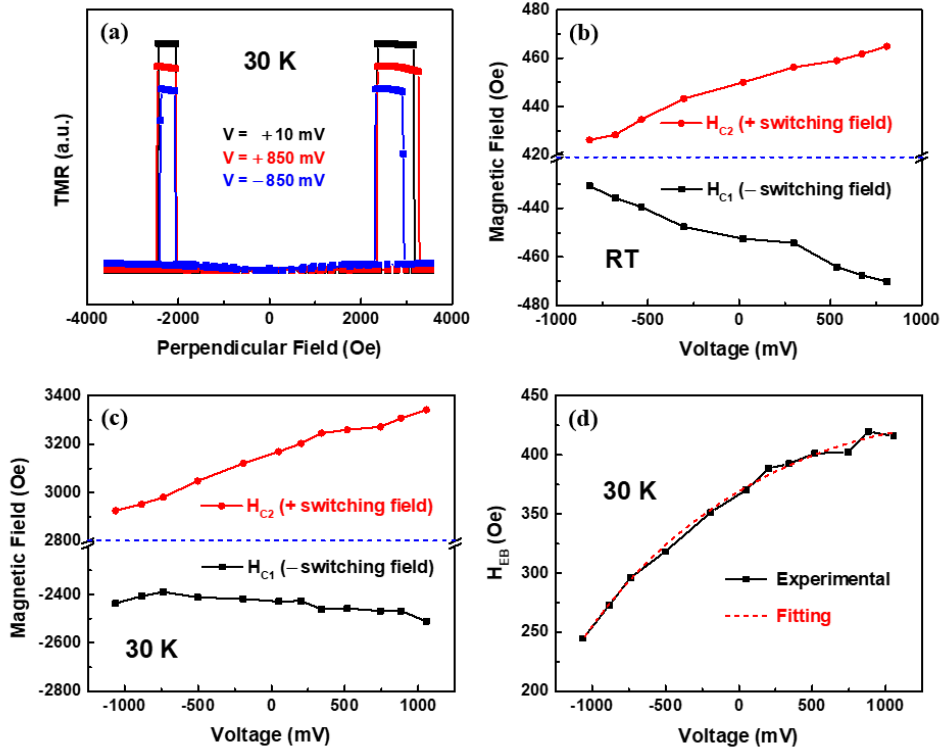

**Fig. 4** (a), Representative TMR curves measured under different bias voltages at 30 K. (b), The voltage dependence of magnetization switching fields $H_{C1}$ and $H_{C2}$ at RT, as a result of the VCMA effect. (c), The voltage dependence of $H_{C1}$ and $H_{C2}$ at 30K, as a result of both the VCMA and voltage-controlled exchange bias effects. (d), Voltage dependence of the exchange bias field at 30K. Experimental data are fitted using equation (1) assuming a linear voltage dependence for $K_{AF}$.

We attempt to understand the voltage dependence of $H_{EB}$ through the model developed by Malozemoff considering the random field at the interface [48],

$$H_{EB} = \frac{2f_i}{\pi^2}\sqrt{A_{AF}K_{AF}}/M_{FM}t_{FM} \qquad (1)$$

where $f_i$ is of order unity and determined by the microscopic interfacial conditions, $A_{AF}$ and $K_{AF}$ are the exchange stiffness and anisotropy of AFM. $H_{EB}$ calculated using this equation is of the same order as experimental observation for typical FM/AFM systems. In Equation 1, $H_{EB}$ depends explicitly on $A_{AF}$, $K_{AF}$, and $M_{FM}$, for a given FM with a fixed thickness. Generally, the dependence of $M_{FM}$ on voltage is negligibly small without the ionic effect [10–12]. $A_{AF}$ can in principle be modified by voltage, but this effect is not considered here because the change of stiffness is also much smaller compared to that of anisotropy as demonstrated in the FM case [49]. The results in Fig. 4d then indicate the change of $H_{EB}$ is due to the control of $K_{AF}$ by voltage. If we assume the AFM anisotropy to be linearly depended on voltage as $K_{AF} = K_0 + \xi * V$, the data indeed can be well fitted by Equation 1 as shown by the red curve in Fig. 4d. The increase in $H_{EB}$ at +1V (+330 mV/nm) corresponds to a 36% increase in $K_{AF}$ compared to the zero-bias value. This behavior of $H_{EB}$ can be well captured by the square root dependence of $K_{AF}$ that is linearly modulated by voltage, which clearly demonstrates the voltage-controlled antiferromagnetism effect in the CoFeB-MgO pMTJs.

Using an empirical $A_{AF}$ value ($10^{-12}$~$10^{-13}$ J/m$^3$) [50], we obtain the effective $K_{AF}$ ~ $10^4$ J/m$^3$, which is about 2 orders of magnitude smaller than $K_{AF}$ of thick AFM layers [51]. From the fitting result of $\xi$ in Fig. 4d, we can extract the linear coefficient of voltage-controlled AF anisotropy to be 22 fJ/Vm (assuming an $A_{AF}$ of $5 \times 10^{-13}$ J/m and Fe(Co)O$_x$ monolayer thickness of 4.3 Å), which represents the scale of voltage-controlled AF effect in this system. The physical origin of the voltage control of $K_{AF}$ is likely related to the electrically induced orbital reconstruction [52]. However, a more quantitative understanding requires further theoretical and experimental investigation, which is beyond the scope of the current work. Here we provide a simplified picture to understand the results in Figure 4: for pinned uncompensated spins below $T_B$, their stability is determined by the interactions with the FM and AF portion of the system. Voltage application may modulate both interactions via voltage-controlled FM and AF anisotropy. In this study, the applied field and the unidirectional magnetic anisotropy (defined as $H_{EB} \times M_{FM} \times t_{FM}$) direction are collinear. Therefore H$_{EB}$ is independent of FM uniaxial anisotropy [53]. Voltage only significantly changes the exchange interaction between the pinned spins and the antiferromagnet by modulating the AF anisotropy, which effectively modifies the pinning site stability and leads to a change in H$_{EB}$ following Malozemoff's model.

Finally, we would also like to comment on the implication of our results on the understanding of the magnetoresistance of CoFeB/MgO/CoFeB MTJs, where a large TMR is always preferred. Our study provides an important clue to understand the discrepancy between the predicted (up to 35,000% [54,55]) and the observed (~400% -1000% [33,56,57] ) TMR values in this system. Among many factors that could reduce the experimental TMR ratio, the interfacial oxidation plays an important role as one monolayer of oxide at the Fe/MgO interface can decrease the TMR by more than 10 times as shown by a DFT calculation [58]. However, prior to our study, it is unknown if a very thin FeO/CoO layer could exist in high-quality pMTJs without detrimentally destroying TMR. No actual TMR has been demonstrated in previous studies where the interfacial oxide was probed by MSHG [42] or ferromagnetic resonance [43]. Here the observation of exchange bias in our high-quality pMTJ with TMR over 200 % (Supplementary Fig. S5) and the TEM-EELS spectrum at CoFeB/MgO interface (Supplementary Fig. S6) unambiguously demonstrates the existence of the FeO/CoO layers. Therefore the interfacial oxidation could be the main reason that limits the TMR in CoFeB/MgO/CoFeB MTJs.

To conclude, we have demonstrated that the exchange bias in the CoFeB/MgO/CoFeB pMTJ with giant magnetoresistance can be effectively controlled by voltage. The square root dependence of the exchange bias field on voltage can be well explained by an AF anisotropy energy that is linearly modulated by voltage. Similar manipulation of exchange bias and antiferromagnetism may be realized in a wide range of other FM/oxide and FM/AFM systems. These results also provide insight into a better understanding of the complex CoFeB/MgO interface and represent a new route to manipulate pMTJs by controlling exchange bias.

**Acknowledgements**

The authors would like to thank Shufeng Zhang and Yihong Cheng for inspiring discussion. This work was supported in part by DARPA through the ERI program (FRANC), by NSF through ECCS-1554011, and by Semiconductor Research through the Global Research Collaboration program. Wei Zhang acknowledges support from AFOSR under Grant no. FA9550-19-1-0254.